\documentclass[superscriptaddress,twocolumn,showpacs,prl,amsmath]{revtex4}
\usepackage{times}
\usepackage{graphicx}
\usepackage{longtable}
\usepackage{color}
\usepackage{amsmath}
\usepackage{amssymb,srcltx}

\newcommand{\beq}{\begin{equation}}
\newcommand{\eeq}{\end{equation}}
\newcommand{\ket} [1] {\vert#1\rangle}
\newcommand{\bra} [1] {\langle#1\vert}

\newcommand{\mean}[1]{\langle #1 \rangle}
\newcommand{\dd}{\text{d}}

\begin{document}

\title{Impact of turbulence in long range quantum and classical communications}

\author{Ivan Capraro}
\affiliation{Department of Information Engineering, University of Padova, via Gradenigo 6/B}
\author{Andrea Tomaello}
\affiliation{Department of Information Engineering, University of Padova, via Gradenigo 6/B}
\author{Alberto Dall'Arche}
\affiliation{Department of Information Engineering, University of Padova, via Gradenigo 6/B}
\author{Francesca Gerlin}
\affiliation{Department of Information Engineering, University of Padova, via Gradenigo 6/B}
\author{Ruper Ursin}
\affiliation{Institute for Quantum Optics and Quantum Information (IQOQI),
Austrian Academy of Sciences, Boltzmanngasse 3, A-1090 Vienna, Austria}
\author{Giuseppe Vallone}
\affiliation{Department of Information Engineering, University of Padova, via Gradenigo 6/B}
\author{Paolo Villoresi}
\affiliation{Department of Information Engineering, University of Padova, via Gradenigo 6/B}

\date{\today}

%%%%%%%%%%%%%%%%%%%%%%%%%%%%%%%%%%%%%%%%%%%%%%%%%%%%%%%%%%%%%%%%%%%

\begin{abstract}
The study of the free-space distribution of quantum correlations is necessary for any future application of quantum as classical communication
 aiming to connect two remote locations. Here we study the propagation of a coherent laser beam
 over 143 Km (between Tenerife and La Palma Islands of the Canary archipelagos). By attenuating the beam we also studied 
the propagation at the single photon level. We investigated the statistic of arrival of the incoming photons and the scintillation of the beam.
From the analysis of the data, we propose the exploitation of turbulence to improve the SNR of the signal.
\end{abstract}

%%%%%%%%%%%%%%%%%%%%%%%%%%%%%%%%%%%%%%%%%%%%%%%%%%%%%%%%%%%%%%%%%%%



%%%%%%%%%%%%%%%%%%%%%%%%%%%%%%%%%%%%%%%%%%%%%%%%%%%%%%%%%%%%%

\maketitle

{\it Introduction - }
The study of the free-space propagation of quantum correlations is necessary for any future application of quantum communication
aiming to connect two remote locations.
The problem related to the free-space propagation is represented by the atmospherical turbulence, that acts as a temporal and spatial 
variation of the air refraction index. A turbulent channel acts an increment of the
 losses on the transmitted photons due to beam-wandering of the beam-centroid or scintillation, increasing the role of the noise \cite{tata61book,fant75ieee,fant75ieee2,fant80ieee,dios04aop}.
 The understanding of the propagation effects induced by turbulence at the receiver as well as the temporal statistics of the 
incoming photons is crucial to assess the quality of the communication and eventually the feasibility of the free-space 
ground-ground and space-ground links \cite{vill08njp,bona09njp,meye11pra}. 

In this work we study the propagation of a free space optical link (143Km) between Tenerife and La Palma Islands of the Canary archipelagos
\cite{ursi07nap, fedr09nap, sche10pnas}. 
The transmitter is located at La Palma, on the roof of the Jacobus Kapteyn Telescope (JKT) and the receiver is the
Optical Ground Station (OGS) at Tenerife. The campaign was performed during the nights between 17 and 25 september 2011.

\begin{figure}[htbp]
  \begin{center}
 \includegraphics[width=8cm]{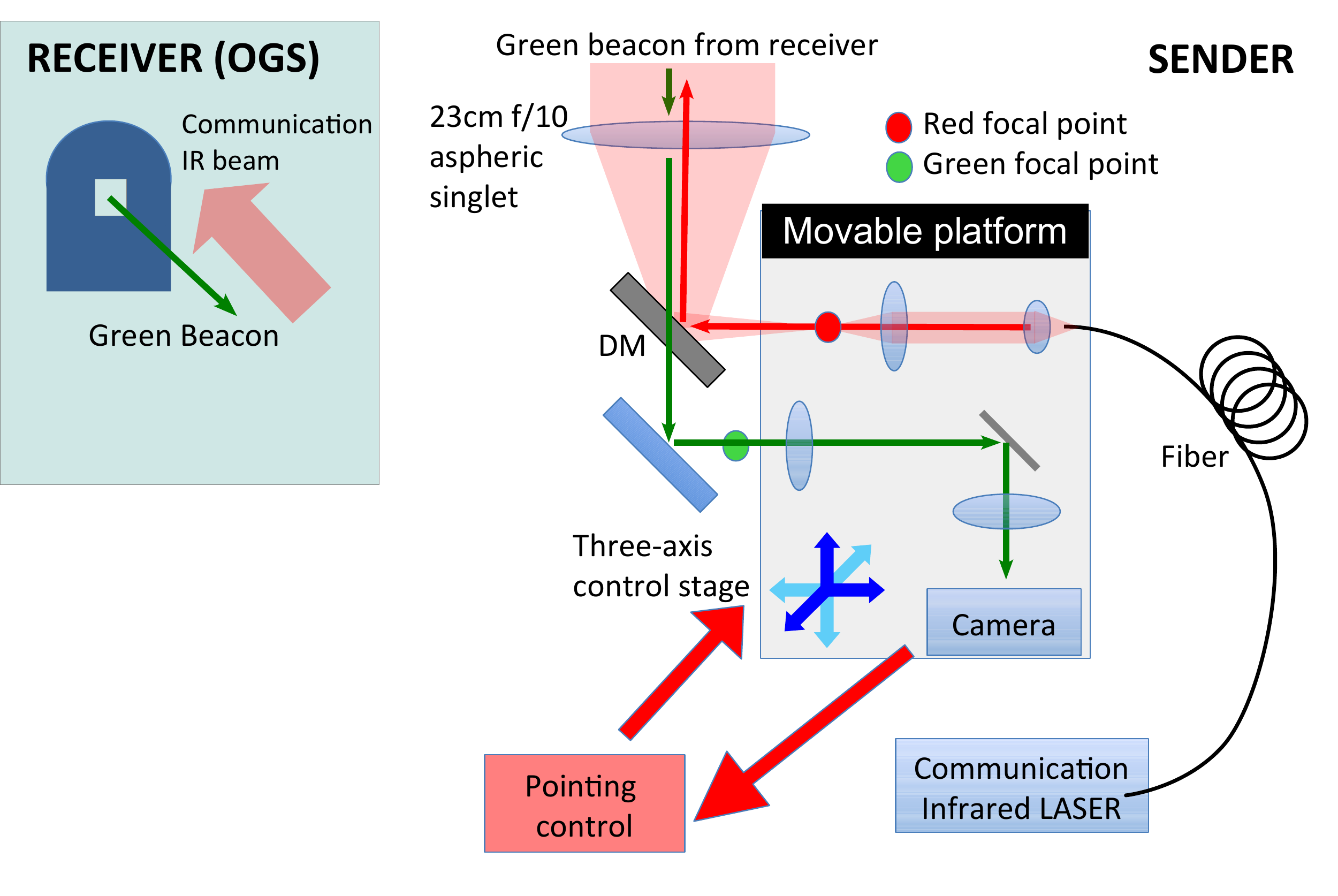}
\caption{\label{fig:setup}Schematic of the optical setup.}
\end{center}
\end{figure}

{\it Optical Setup - }
The optical setup of the transmitter is shown in figure \ref{fig:setup}. 
It consists of suitably designed telescope whose key component is a singlet aspheric lens 
of 23 cm diameter and 220 cm focal lenght at $810$nm. The lens diameter was chosen to be significantly greater of the estimated
Fried parameter $r_0$ \cite{frie66josa} in order to obtain a spot at OGS compared
 to the telescope primary mirror and consequently a greater power transfer between the two sites. 
Our light source is an infrared diode at 808nm coupled into single mode fiber with output power of about 6mW and suitable attenuators. 
In order to facilitate the raw pointing a mechanical XY stage has been added (we define the Z direction as the optical axis of the system). 
This stage moves all the 2.5m long telescope in the XY plane.
All the structure is assembled by three aluminum flanges,  
one hods the lens, one the focal plane and the other is attached to the XY back stage.
The lens is fixed to a articulated mount to prevent bending of the structure.

The IR source has been aligned by means of a dichroic mirror (DM) reflecting the 808nm radiation.
The platform carrying the focusing lens, the collimating lens and the fiber port for the IR can be moved by a 
micro metric XYZ system based on stepped motors. 
In this way the beam can be slightly steared by moving the focal spot at the singlet focus.
The instantaneous deviation from the initial pointing direction is acquired by using a 532nm beacon beam sent from the receiver
by using a small portable low power laser module directly pointed towards La Palma without any optics.  
The beacon laser is acquired with a CMOS camera placed on the movable platform after  the DM trasmission. 
The centroid of the beacon spot on the camera determines the 
correction on the outcoming IR laser by means of an error signal with respect to the reference position. 
The position of the spot at the camera are sampled about once a second and averaged for a number of frames and this data feeds a control
 software that calculates the movement for the fine XY stage in order to compensate slow drift in beam direction,
 which was performed in 1 sec timescale.

We collected data at the OGS in Tenerife in order to measure
the received power, the scintillation and analyze the temporal structure of the signal. 
We placed in the OGS Coud\`e focus a photodiode, a power meter and, when 
the beam is suitably attenuated with neutral filter, we also collected data with a single photon (Excelitas SPCM-AQRH model) detector (SPAD). 
The following data were recorded during the nights between 21th and the 24th September 2011. 
We obtained an average attenuation of about 30dB for many times during the best run with peaks of 27dB averaged over 2 minutes. 
The attenuation is calculated from the fiber and not from the singlet lens: the attenuation of the telescope is thus included in the measured attenuation.

\begin{figure}[t]
 \includegraphics[width=8cm]{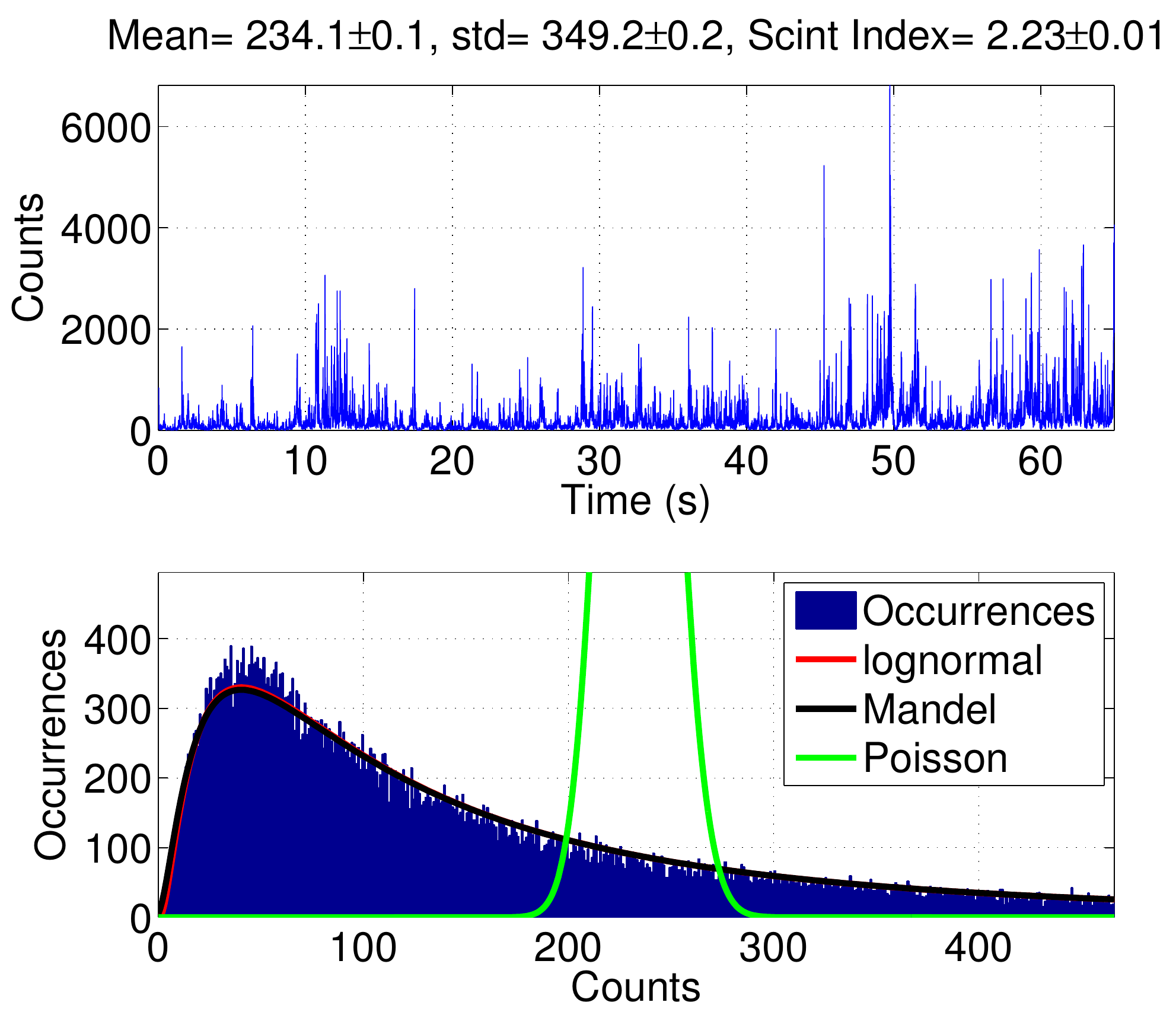}
\caption{{SPAD temporal distribution  of count occurences and corresponding lognormal and Mandel curves. 
We can compare the data with the corresponding Poissonian distribution with the same mean value ($234.1\pm0.1$) 
that would be obtained without turbolence.}
}
\label{fig:ST}
\end{figure}

{\it Link Analisys - } Let's first describe  the single photon detection acquisition. 
We performed several measurements by setting the counting interval $T$ to $0.1ms$, $1ms$ and $10ms$.
Due to turbulence effects,  the mean photon number $q$ in a counting interval at the receiver 
should follow a lognormal probability distribution \cite{milo04job}:
\beq
\label{lognormal}
P(q)=\frac{1}{q\sqrt{2\pi \sigma^2}}e^{-[(\ln \frac{q}{\langle q\rangle}+\frac12\sigma^2)]^2/(2\sigma^2)}
\eeq 
where $\mean q$ is the average, $\sigma^2=\ln(1+SI)$  and $SI=\frac{\Delta q^2}{\langle q\rangle^2}$ is the scintillation index. 
If the counting interval $T$ is large compared with the coherence time of the source 
and $T$ is short compared with the turbulence timescale, the probability of detecting $n$ photon in each 
interval follow the Mandel distribution:
\beq
p_n=\int\dd q\;\frac{q^ne^{-q}}{n!}P(q)
\eeq 
Note that the mean number of detected photon is
 $\langle n\rangle=\sum_nnp_n=\langle q\rangle$.
We report the analysis of the temporal distribution {of an acquisition with $1ms$ counting interval}  in figure \ref{fig:ST}(top).
{It is possible to observe that, when the average number of detected photons, $\langle n\rangle$, is large 
(typically larger than 50) and the scintillation index bigger than 1,
the lognormal and Mandel distribution are quite similar.}
{Given  the experimental scintillation as $2.23\pm0.01$ and the mean value of detected photon as 234, we  
show the counting occurrences together with the corresponding lognormal distribution in figure \ref{fig:ST}(bottom). For comparison, 
we also insert the corresponding Mandel distribution with the $\langle q\rangle=234.1\pm0.1$ and $\sigma=349.2\pm0.2$ parameter obtained
from the raw data. 
We evaluated the similarity between the experimental data and the lognormal or the Mandel curve, defined as
$S=\frac{(\sum\sqrt{p_iq_i})^2}{\sum p_i\sum q_j}$ where $p_i$ and $q_i$ are respectively the theoretical and experimental
occurrence. The similarity of the lognormal curve with the data is 0.9959 while the Mandel curve has a similarity of 0.9967
showing a clear evidence of the statistic transformation.}
{The green curve represents the corresponding Poisson distribution with the same observed mean value,
to compare what it have been obtained if the statistic of arrival photon would be purely poissoninan.
}
In table \ref{table:SPAD_data} we report the data obtained for several different SPAD acquisitions.

\begin{figure}[h]
  \begin{center}
 \includegraphics[width=8cm]{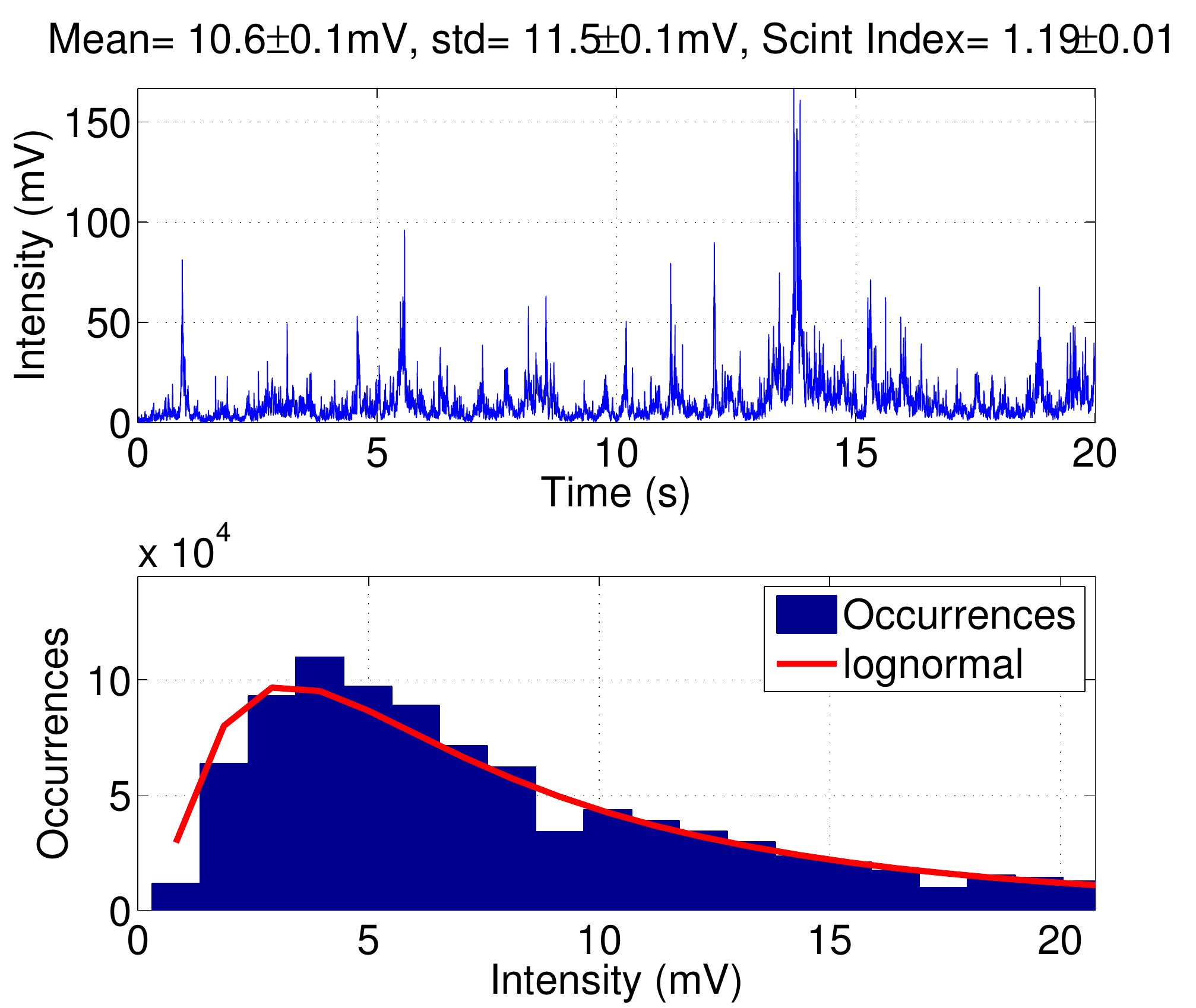} 
\caption{Photodiode temporal distribution intensity occurences and corresponding  lognormal curve.}
\label{fig:temp_ph}
\end{center}
\end{figure}

As said, we also measure the intensity of received light with a fast photodiode by using an intense laser source.
In figure \ref{fig:temp_ph} we plot the {temporal distribution of the photodiode voltage} of a data set covering 20 s.
The intensities are recorded with 50kHz frequency. 
{Also in this case the intensity occurences follows a lognormal distribution \eqref{lognormal} as shown from the lognormal curve with
similarity of 0.9896.
In this case the scintillation index evaluated from the experimental data is $SI=\frac{\Delta I^2}{\langle I\rangle^2}=1.19\pm0.01$.}

{\it Improving the SNR - }
For both quantum and classical communication, it is of paramount importance achieving an high
signal to noise ratio (SNR). If a qubit state $\ket{\psi}$ {encoded in the photon polarization} must be sent between
two remote location, 
it is possible to determine the effect of (white) noise on the {polarization} fidelity\footnote{{If $\ket{\psi}$
is the polarization state of the sent photon and $\rho$ is the polarization density matrix of the received photon, the fidelity
$F=\bra{\phi}\rho\ket{\phi}$ is the probability of receiving the correct state.}}. Let's measure the SNR
in dB, namely SNR$=10\log_{10}\frac{N_s}{N_n}$, 
where $N_n$ the average amount of noise
(coming from dark detections or background radiation) and $N_s$ is the average counting signal.
It is easy to show that the fidelity depends on the SNR as:
\beq
F=1-\frac{1}{1+10^{SNR/10}}\,.
\eeq

In order to improve the SNR for the transmission of single photons in a long distance free-space link as the present one which
use a 1m optical receiver, out of our findings we envisage the exploitation of the following procedure.
With a given frequency (slower than the single photon {transmission rate}) the free-space channel
is probed by means of a classical signal that gives the information of the instantaneous transmission of the channel.
Only if the transmission is above a given threshold the single photon signal is acquired. It is crucial for the protocol to be
efficient to correctly identify the "probing" frequency and the threshold to be used. 
This technique can be also used in the classical case, for instance in the on-off keying.

\begin{figure}[t]
 \includegraphics[width=8cm]{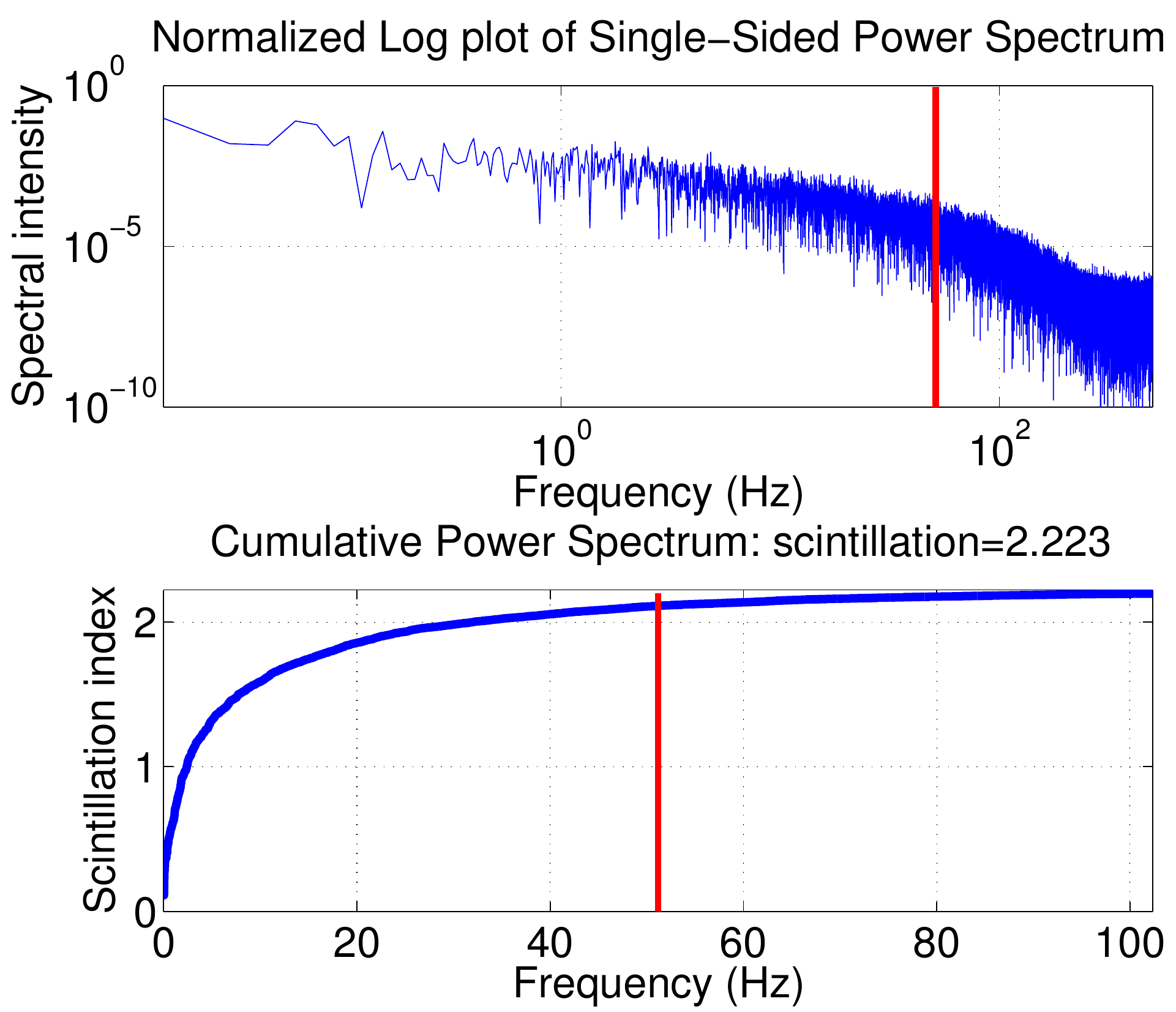}
\caption{ SPAD power spectrum and cumulative power spectrum. 
 Frequency bound (red line): the frequencies below 51.1724 Hz contribute to the 95\% of the scintillation index. 
}
\label{fig:STspectrum}
\end{figure}
We report in figure
\ref{fig:STspectrum} the frequency spectrum and the cumulative power spectrum of the data plotted in figure \ref{fig:ST}. 
The normalized plot of the power spectrum is
obtained by normalizing the intensities by the average $I'=I/\langle I\rangle$.
The power spectrum is related to the scintillation index as follow. We write the set of (normalized) acquisitions as $I'_k$ 
with $k=0,\cdots, N-1$ and $N=20s/20\mu s=10^6$ 
the number of intensity acquisitions over the 20 seconds. The Fourier components are given by $\widetilde I_n=\sum_k I'_k\omega^{nk}$ with
$\omega=e^{-\frac{2\pi i}{N}}$. By Parseval's theorem it is easy to show that $SI=\frac{2}{N^2}\sum^{N/2}_{n=1}|\widetilde I_n|^2$, namely
it is the cumulative power spectrum without the zero frequency ($\widetilde I_0$) component.
We can notice that the frequencies  contributing to the scintillation (up to 95$\%$) are within (almost) 50Hz. For frequencies above
around 500Hz, the spectrum becomes flat, indicating that at this frequency the random noise is dominant.
The typical fluctuations of the transmission channel due to turbulence are within 100Hz (see table \ref{table:SPAD_data}).
The frequency analysis of the temporal scintillation indicates that the probing frequency doesn't need to be higher than 1KHz.
\begin{figure}[t]
 \includegraphics[width=8cm]{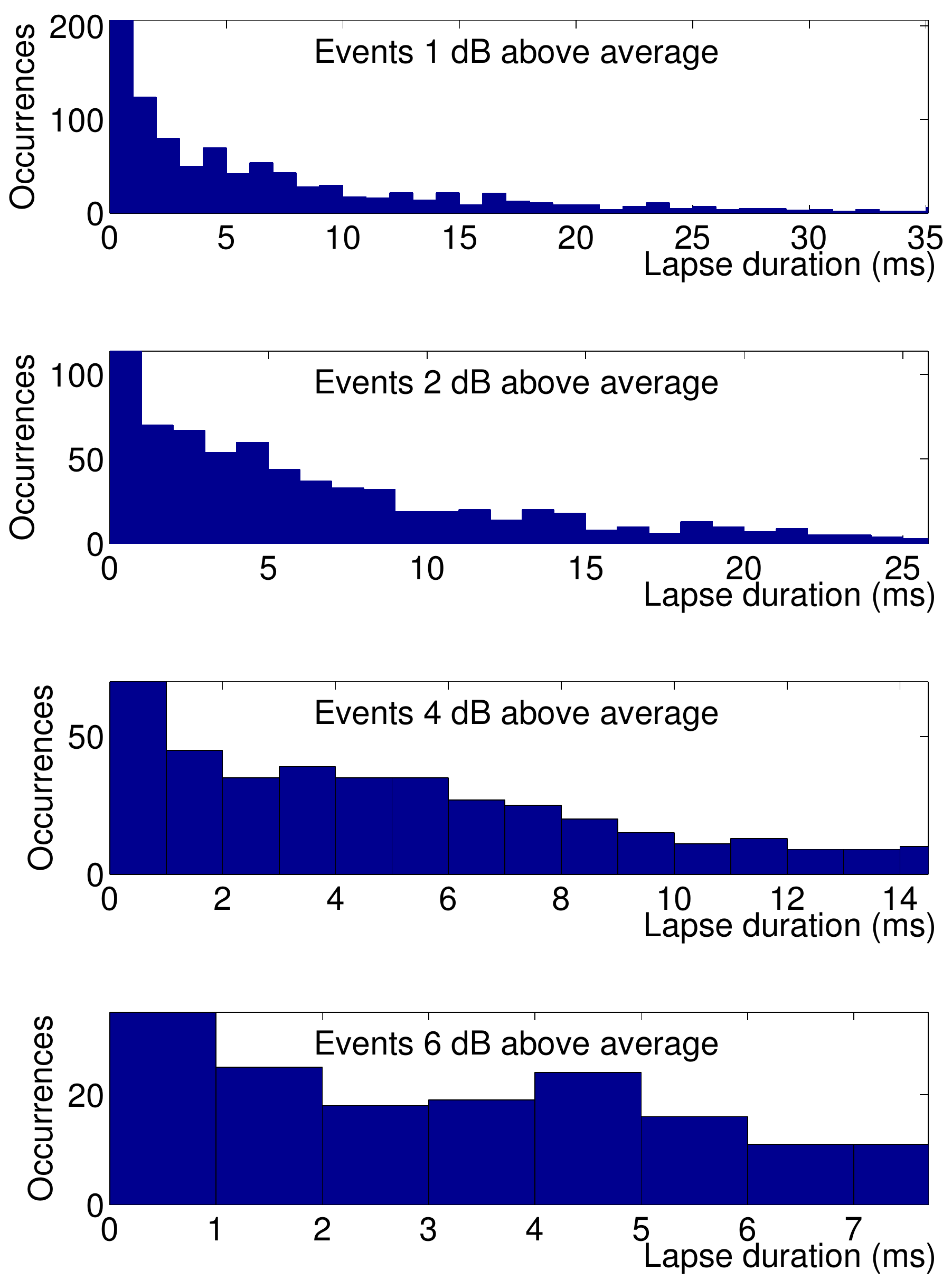}
\caption{Duration (in ms) of events with over-threshold counting: in the different plots
we considered a threshold of 1, 2, 4 and 6 dB above the average.}
\label{fig:ffs}
\end{figure}

\begin{figure}[t]
 \includegraphics[width=7cm]{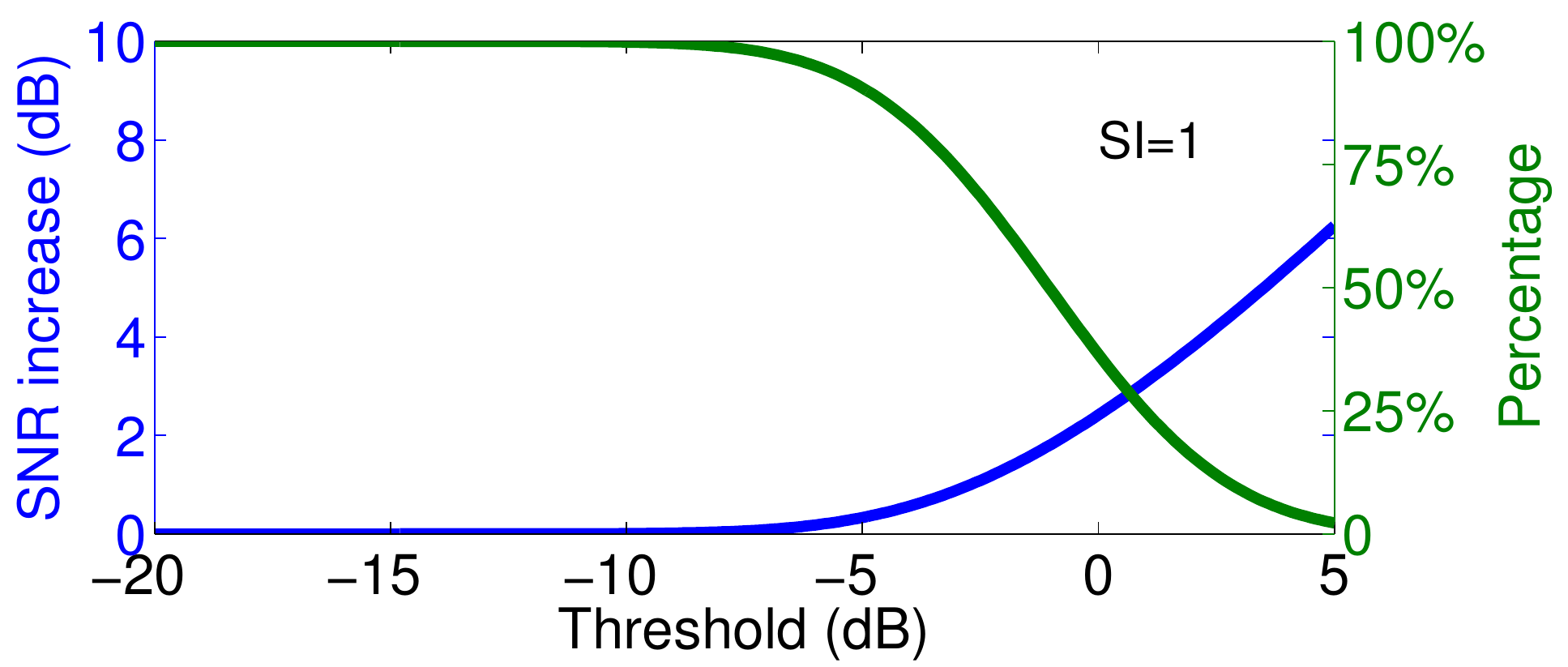}
\\
 \includegraphics[width=7cm]{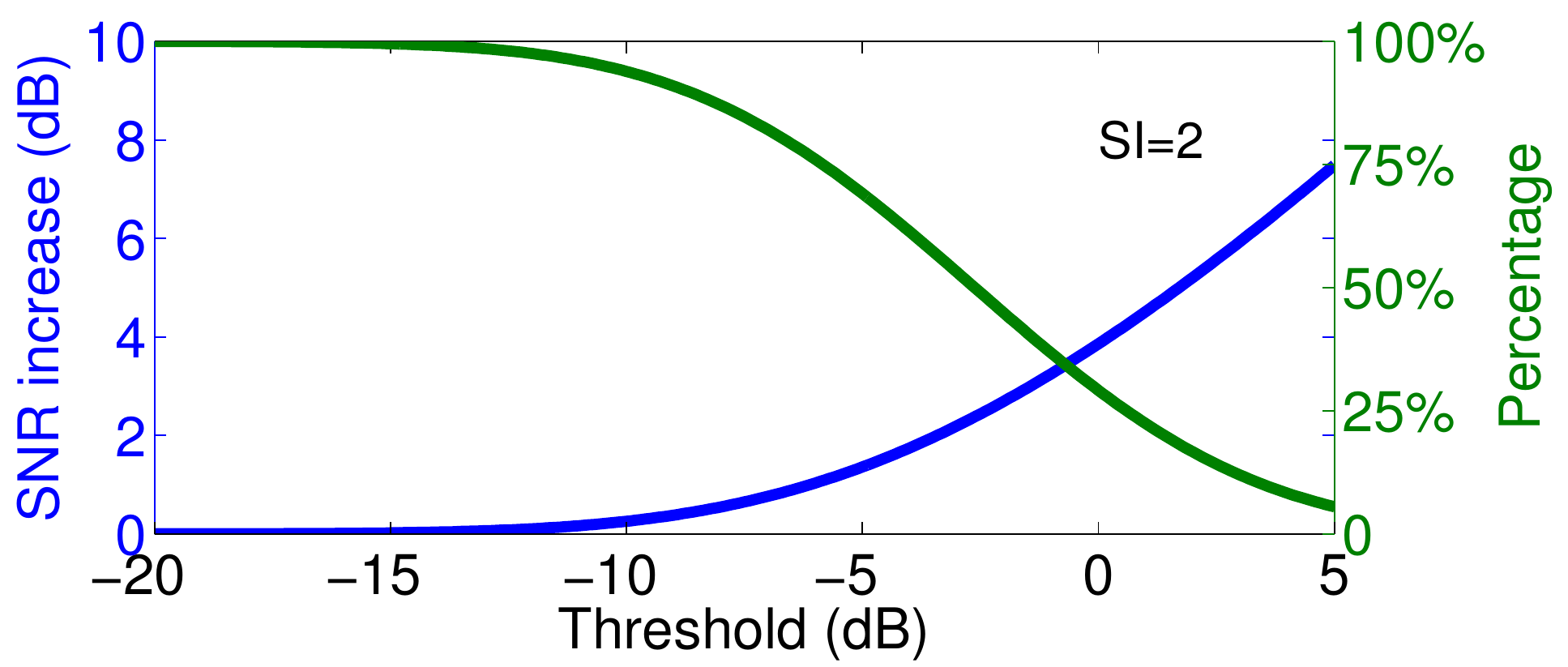}
\\
 \includegraphics[width=7cm]{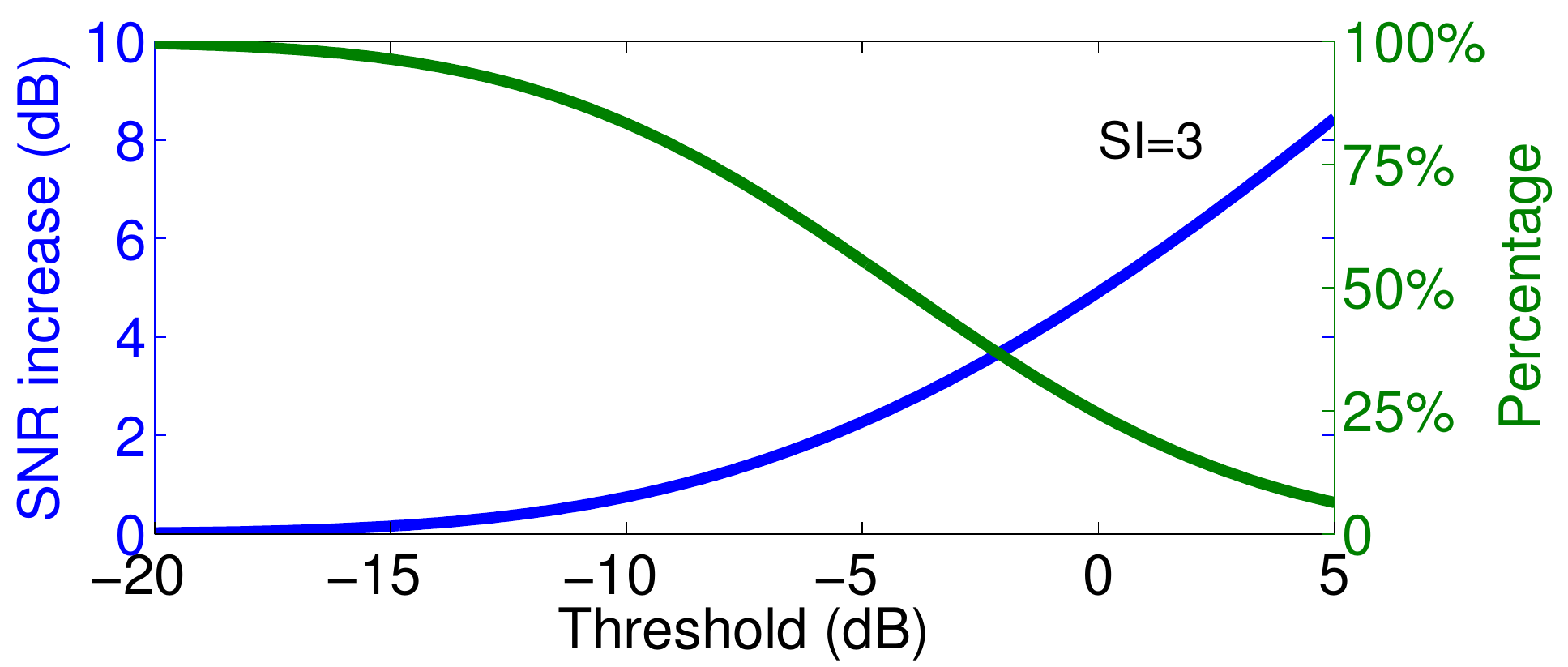}
\caption{SNR and the percentage
of the overall counts that will be detected in function of the threshold selection.}
\label{fig:threshold}
\end{figure}

In order to obtain a further evidence, we analysed the features of the counts above a given threshold of the 
signal reported in fig. \ref{fig:ST}. By considering a threshold of 1, 2, 4 and 6 dB above the average
we considered the duration (in ms) of events with over-threshold counting. The results are shown in figure
\ref{fig:ffs}. The probability of obtaining an event above a given threshold $q_0$
 can be predicted from the lognormal distribution\footnote{Here we replaced
the Mandel distribution with the lognormal distribution.}
\beq
p(q>q_0)=\frac12-\frac12\text{erf}[\frac{\ln \frac{q_0}{\langle q\rangle}+\frac12\sigma^2}{\sqrt{2\sigma^2}}]
\eeq
where erf$(x)$ is the Gaussian error function erf$(x)=\frac{2}{\sqrt\pi}\int^x_0e^{-t^2}dt$.
Acquiring the single photon channel only if the "probed" transmission is above a given threshold implies an increase
of the average photon counts in each time-slot.
It is possible to show that, by considering only the events in which the transmission satisfy $T>T_0$,
 the new mean value $\langle n\rangle_{thres}$ is
\beq
\frac{\langle n\rangle_{thres}}{\langle n\rangle }=
\frac{1-\text{erf}[\frac{\ln \frac{T_0}{\langle T\rangle}-\frac12\sigma^2}{\sqrt{2\sigma^2}}]}
{1-\text{erf}[\frac{\ln \frac{T_0}{\langle T\rangle}+\frac12\sigma^2}{\sqrt{2\sigma^2}}]}>1
\eeq
Clearly, this threshold selection increase the SNR, but at the same time decrease the overall
counts in given time. In figure \ref{fig:threshold} we show the increase (in dB) of the SNR and the percentage
of the overall counts that will be detected.
In cases of strong turbulence and high noise, this technique could help
the qubit transmission by "exploiting turbulence", namely considering only the particular moments
in which the turbulence increase the channel transmission.

In conclusion, the statistic of arrival of single photon over free-space 143Km optical link has been analyzed demonstrating
the transformation from Poissonian to Lognormal distribution thus expanding this investigation 
for more than an order of magnitude in length with respect to previous results \cite{milo04job}.
The evidence of  consecutive subintervals of low losses allows to  envisage the exploitation of turbulence as an SNR improvement
technique.

\begin{acknowledgments} 
The authors wish to thank the staff of IAC: F. Sanchez-Martinez, A. Alonso, C. Warden, M. Serra, J. Carlos and the staff of ING: 
M. Balcells, C. Benn, J. Rey, O. Vaduvescu, A. Chopping, D. Gonza\'ez, S. Rodr\'iguez, M. Abreu, L. Gonza\'ez; J. Kuusela, E. Wille, as well as
Z. Sodnik, and J. Perdigues of the OGS of ESA.
This work has been carried out within the Strategic-Research-Project QUINTET of the Department of
Information Engineering, University of Padova and the
Strategic-Research-Project QUANTUMFUTURE of the University of
Padova.
\end{acknowledgments}

\begin{table}
\begin{tabular}{|l|l|l|l|l|l|l|l|}
&Mean&Std&Time(s)&{Window} (ms)&SI&Bound(Hz)\\\hline
\hline
%23-24background-1ms-screen.mat
backgr.
& 0.4732&0.7256&10&1&2.3515&475\\\hline
%23-24HiSignal10ms.mat
Hi
&5291&9135&650&10&2.9805&22\\\hline
%23-24HiSignal1ms-1.mat
Hi
&678.7&820.8&65&1&1.4626&43\\\hline
%23-24HiSignal1ms-2.mat
Hi
&510.4&751.9&65&1&2.1704&45\\\hline
%23-24HiSignal1ms-3.mat
Hi
&781.5&951.9&65&1&1.4834&42\\\hline
%23-24HiSignal1ms-4.mat
Hi
&180.2&312.0&65&1&2.997&39\\\hline
%23-24HiSignal1ms-5.mat
Hi
&234.1&349.2&65&1&2.2251&51\\\hline
%23-24HiSignal100us-1.mat
Hi
&43.18&48.76&6.5&0.1&1.2748&81\\\hline
%23-24HiSignal100us-2.mat
Hi
&21.17&26.34&6.5&0.1&1.5485&88\\\hline
%23-24HiSignal100us-3.mat
Hi
&75.37&132.10&6.5&0.1&3.072&35\\\hline
%23-24HiSignal100us-4.mat
Hi
&59.97&105.01&6.5&0.1&3.0659&37\\\hline
%23-24LoSignal10ms-1.mat
Low
&37.86&48.42&200&10&1.6355&35\\\hline
%23-24LoSignal10ms-2.mat
Low
&20.83&24.18&200&10&1.3479&35\\\hline
%23-24LoSignal1ms-1.mat
Low
&2.862&3.795&65&1&1.7578&384\\\hline
%23-24LoSignal1ms-2.mat
Low
&5.267&7.519&65&1&2.0383&258\\\hline
\end{tabular}
\caption{Data obtained for different single photon acquisition compared to the background (first line). 
For each acquisition we report the total duration of the acquisition (Time),
{the temporal windows defining the counting interval (Window), 
the mean number of counts in the counting interval (Mean)} and its the standard deviation (Std). We also report the
scintillation index (SI) and {the frequency bound such that all the frequencies below the bound  
contribute up to 95\% of the scintillation (Bound)}. With High (Low) we indicate acquisition
with high (low) mean photon number detected during 1s. We notice that for the last two data sets the bound is higher due to the 
low signal compared to the background (having flat frequency spectrum). }
\label{table:SPAD_data}
\end{table}

%%%%%%%%%%%%%%%%%%%%%%%%%%%%%%%%%%%%%%%%%%%%%%%%%%%%%%%%%%%%%
%%%%% References %%%%%

%\bibliography{../../../bibliografia/library}% Produces the bibliography via BibTeX.

\begin{thebibliography}{13}
\expandafter\ifx\csname natexlab\endcsname\relax\def\natexlab#1{#1}\fi
\expandafter\ifx\csname bibnamefont\endcsname\relax
  \def\bibnamefont#1{#1}\fi
\expandafter\ifx\csname bibfnamefont\endcsname\relax
  \def\bibfnamefont#1{#1}\fi
\expandafter\ifx\csname citenamefont\endcsname\relax
  \def\citenamefont#1{#1}\fi
\expandafter\ifx\csname url\endcsname\relax
  \def\url#1{\texttt{#1}}\fi
\expandafter\ifx\csname urlprefix\endcsname\relax\def\urlprefix{URL }\fi
\providecommand{\bibinfo}[2]{#2}
\providecommand{\eprint}[2][]{\url{#2}}

\bibitem[{\citenamefont{Tatarski}(1961)}]{tata61book}
\bibinfo{author}{\bibfnamefont{V.~I.} \bibnamefont{Tatarski}},
  \emph{\bibinfo{title}{{Wave Propagation in a Turbulent Medium}}}
  (\bibinfo{publisher}{McGraw-Hill}, \bibinfo{year}{1961}).

\bibitem[{\citenamefont{Fante}(1975{\natexlab{a}})}]{fant75ieee}
\bibinfo{author}{\bibfnamefont{R.~L.} \bibnamefont{Fante}},
  \bibinfo{journal}{Proc. IEEE} \textbf{\bibinfo{volume}{63}},
  \bibinfo{pages}{1669} (\bibinfo{year}{1975}{\natexlab{a}}).

\bibitem[{\citenamefont{Fante}(1975{\natexlab{b}})}]{fant75ieee2}
\bibinfo{author}{\bibfnamefont{R.}~\bibnamefont{Fante}}, \bibinfo{journal}{IEEE
  Trans Antenms Propagat.} \textbf{\bibinfo{volume}{AP-23}},
  \bibinfo{pages}{382} (\bibinfo{year}{1975}{\natexlab{b}}).

\bibitem[{\citenamefont{Fante}(1980)}]{fant80ieee}
\bibinfo{author}{\bibfnamefont{R.~L.} \bibnamefont{Fante}},
  \bibinfo{journal}{Proc. IEEE} \textbf{\bibinfo{volume}{68}},
  \bibinfo{pages}{1424} (\bibinfo{year}{1980}).

\bibitem[{\citenamefont{Dios et~al.}(2004)\citenamefont{Dios, Rubio,
  Rodr\'{\i}guez, and Comer\'{o}n}}]{dios04aop}
\bibinfo{author}{\bibfnamefont{F.}~\bibnamefont{Dios}},
  \bibinfo{author}{\bibfnamefont{J.~A.} \bibnamefont{Rubio}},
  \bibinfo{author}{\bibfnamefont{A.}~\bibnamefont{Rodr\'{\i}guez}},
  \bibnamefont{and}
  \bibinfo{author}{\bibfnamefont{A.}~\bibnamefont{Comer\'{o}n}},
  \bibinfo{journal}{Applied optics} \textbf{\bibinfo{volume}{43}},
  \bibinfo{pages}{3866} (\bibinfo{year}{2004}), ISSN \bibinfo{issn}{0003-6935}.

\bibitem[{\citenamefont{Villoresi et~al.}(2008)\citenamefont{Villoresi,
  Jennewein, Tamburini, Aspelmeyer, Bonato, Ursin, Pernechele, Luceri, Bianco,
  Zeilinger et~al.}}]{vill08njp}
\bibinfo{author}{\bibfnamefont{P.}~\bibnamefont{Villoresi}},
  \bibinfo{author}{\bibfnamefont{T.}~\bibnamefont{Jennewein}},
  \bibinfo{author}{\bibfnamefont{F.}~\bibnamefont{Tamburini}},
  \bibinfo{author}{\bibfnamefont{M.}~\bibnamefont{Aspelmeyer}},
  \bibinfo{author}{\bibfnamefont{C.}~\bibnamefont{Bonato}},
  \bibinfo{author}{\bibfnamefont{R.}~\bibnamefont{Ursin}},
  \bibinfo{author}{\bibfnamefont{C.}~\bibnamefont{Pernechele}},
  \bibinfo{author}{\bibfnamefont{V.}~\bibnamefont{Luceri}},
  \bibinfo{author}{\bibfnamefont{G.}~\bibnamefont{Bianco}},
  \bibinfo{author}{\bibfnamefont{A.}~\bibnamefont{Zeilinger}},
  \bibnamefont{et~al.}, \bibinfo{journal}{New Journal of Physics}
  \textbf{\bibinfo{volume}{10}}, \bibinfo{pages}{033038}
  (\bibinfo{year}{2008}), ISSN \bibinfo{issn}{1367-2630}.

\bibitem[{\citenamefont{Bonato et~al.}(2009)\citenamefont{Bonato, Tomaello, {Da
  Deppo}, Naletto, and Villoresi}}]{bona09njp}
\bibinfo{author}{\bibfnamefont{C.}~\bibnamefont{Bonato}},
  \bibinfo{author}{\bibfnamefont{A.}~\bibnamefont{Tomaello}},
  \bibinfo{author}{\bibfnamefont{V.}~\bibnamefont{{Da Deppo}}},
  \bibinfo{author}{\bibfnamefont{G.}~\bibnamefont{Naletto}}, \bibnamefont{and}
  \bibinfo{author}{\bibfnamefont{P.}~\bibnamefont{Villoresi}},
  \bibinfo{journal}{New Journal of Physics} \textbf{\bibinfo{volume}{11}},
  \bibinfo{pages}{045017} (\bibinfo{year}{2009}), ISSN
  \bibinfo{issn}{1367-2630}.

\bibitem[{\citenamefont{Meyer-Scott et~al.}(2011)\citenamefont{Meyer-Scott,
  Yan, MacDonald, Bourgoin, H\"{u}bel, and Jennewein}}]{meye11pra}
\bibinfo{author}{\bibfnamefont{E.}~\bibnamefont{Meyer-Scott}},
  \bibinfo{author}{\bibfnamefont{Z.}~\bibnamefont{Yan}},
  \bibinfo{author}{\bibfnamefont{A.}~\bibnamefont{MacDonald}},
  \bibinfo{author}{\bibfnamefont{J.-P.} \bibnamefont{Bourgoin}},
  \bibinfo{author}{\bibfnamefont{H.}~\bibnamefont{H\"{u}bel}},
  \bibnamefont{and}
  \bibinfo{author}{\bibfnamefont{T.}~\bibnamefont{Jennewein}},
  \bibinfo{journal}{Physical Review A} \textbf{\bibinfo{volume}{84}},
  \bibinfo{pages}{1} (\bibinfo{year}{2011}), ISSN \bibinfo{issn}{1050-2947}.

\bibitem[{\citenamefont{Ursin et~al.}(2007)\citenamefont{Ursin, Backus,
  Tiefenbacher, Schmitt-Manderbach, Weier, Scheidl, Lindenthal, Blauensteiner,
  Jennewein, Perdigues et~al.}}]{ursi07nap}
\bibinfo{author}{\bibfnamefont{R.}~\bibnamefont{Ursin}},
  \bibinfo{author}{\bibfnamefont{S.}~\bibnamefont{Backus}},
  \bibinfo{author}{\bibfnamefont{H.~C. K.~F.} \bibnamefont{Tiefenbacher}},
  \bibinfo{author}{\bibfnamefont{T.}~\bibnamefont{Schmitt-Manderbach}},
  \bibinfo{author}{\bibfnamefont{H.}~\bibnamefont{Weier}},
  \bibinfo{author}{\bibfnamefont{T.}~\bibnamefont{Scheidl}},
  \bibinfo{author}{\bibfnamefont{M.}~\bibnamefont{Lindenthal}},
  \bibinfo{author}{\bibfnamefont{B.}~\bibnamefont{Blauensteiner}},
  \bibinfo{author}{\bibfnamefont{T.}~\bibnamefont{Jennewein}},
  \bibinfo{author}{\bibfnamefont{J.}~\bibnamefont{Perdigues}},
  \bibnamefont{et~al.}, \bibinfo{journal}{Nature Physics}
  \textbf{\bibinfo{volume}{3}}, \bibinfo{pages}{481} (\bibinfo{year}{2007}).

\bibitem[{\citenamefont{Fedrizzi et~al.}(2009)\citenamefont{Fedrizzi, Ursin,
  Herbst, Nespoli, Prevedel, Scheidl, Tiefenbacher, Jennewein, and
  Zeilinger}}]{fedr09nap}
\bibinfo{author}{\bibfnamefont{A.}~\bibnamefont{Fedrizzi}},
  \bibinfo{author}{\bibfnamefont{R.}~\bibnamefont{Ursin}},
  \bibinfo{author}{\bibfnamefont{T.}~\bibnamefont{Herbst}},
  \bibinfo{author}{\bibfnamefont{M.}~\bibnamefont{Nespoli}},
  \bibinfo{author}{\bibfnamefont{R.}~\bibnamefont{Prevedel}},
  \bibinfo{author}{\bibfnamefont{T.}~\bibnamefont{Scheidl}},
  \bibinfo{author}{\bibfnamefont{F.}~\bibnamefont{Tiefenbacher}},
  \bibinfo{author}{\bibfnamefont{T.}~\bibnamefont{Jennewein}},
  \bibnamefont{and}
  \bibinfo{author}{\bibfnamefont{A.}~\bibnamefont{Zeilinger}},
  \bibinfo{journal}{Nature Physics} \textbf{\bibinfo{volume}{5}},
  \bibinfo{pages}{389} (\bibinfo{year}{2009}).

\bibitem[{\citenamefont{Scheidl et~al.}(2010)\citenamefont{Scheidl, Ursin,
  Kofler, Ramelow, Ma, Herbst, Ratschbacher, Fedrizzi, Langford, Jennewein
  et~al.}}]{sche10pnas}
\bibinfo{author}{\bibfnamefont{T.}~\bibnamefont{Scheidl}},
  \bibinfo{author}{\bibfnamefont{R.}~\bibnamefont{Ursin}},
  \bibinfo{author}{\bibfnamefont{J.}~\bibnamefont{Kofler}},
  \bibinfo{author}{\bibfnamefont{S.}~\bibnamefont{Ramelow}},
  \bibinfo{author}{\bibfnamefont{X.-S.} \bibnamefont{Ma}},
  \bibinfo{author}{\bibfnamefont{T.}~\bibnamefont{Herbst}},
  \bibinfo{author}{\bibfnamefont{L.}~\bibnamefont{Ratschbacher}},
  \bibinfo{author}{\bibfnamefont{A.}~\bibnamefont{Fedrizzi}},
  \bibinfo{author}{\bibfnamefont{N.~K.} \bibnamefont{Langford}},
  \bibinfo{author}{\bibfnamefont{T.}~\bibnamefont{Jennewein}},
  \bibnamefont{et~al.}, \bibinfo{journal}{Proc. Natl. Acad. Sci. USA}
  \textbf{\bibinfo{volume}{107}}, \bibinfo{pages}{19708}
  (\bibinfo{year}{2010}).

\bibitem[{\citenamefont{Fried}(1966)}]{frie66josa}
\bibinfo{author}{\bibfnamefont{D.~L.} \bibnamefont{Fried}},
  \bibinfo{journal}{Journal of the Optical Society of America}
  \textbf{\bibinfo{volume}{56}}, \bibinfo{pages}{1372} (\bibinfo{year}{1966}).

\bibitem[{\citenamefont{Milonni et~al.}(2004)\citenamefont{Milonni, Carter,
  Peterson, and Hughes}}]{milo04job}
\bibinfo{author}{\bibfnamefont{P.~W.} \bibnamefont{Milonni}},
  \bibinfo{author}{\bibfnamefont{J.~H.} \bibnamefont{Carter}},
  \bibinfo{author}{\bibfnamefont{C.~G.} \bibnamefont{Peterson}},
  \bibnamefont{and} \bibinfo{author}{\bibfnamefont{R.~J.}
  \bibnamefont{Hughes}}, \bibinfo{journal}{Journal of Optics B: Quantum and
  Semiclassical Optics} \textbf{\bibinfo{volume}{6}}, \bibinfo{pages}{S742}
  (\bibinfo{year}{2004}), ISSN \bibinfo{issn}{1464-4266}.

\end{thebibliography}

\end{document}